\documentclass[twocolumn,showpacs,preprintnumbers,superscriptaddress,amsmath,amssymb]{revtex4}

\usepackage{epsfig}
\usepackage{graphicx}
\usepackage{dcolumn}
\usepackage{bm}
\usepackage{times}
\usepackage{xcolor}
\usepackage{multirow}
\usepackage[percent]{overpic}

\begin{document}


\title{Promoting information spreading by using contact memory}

\author{Lei Gao} 
\affiliation{Web Sciences Center, University of Electronic
Science and Technology of China, Chengdu 610054, China}
\affiliation{Big Data Research Center, University of Electronic
Science and Technology of China, Chengdu 610054, China}

\author{Wei Wang}  \email{wwzqbx@hotmail.com}
\affiliation{Web Sciences Center, University of Electronic
Science and Technology of China, Chengdu 610054, China}
\affiliation{Big Data Research Center, University of Electronic
Science and Technology of China, Chengdu 610054, China}

\author{Panpan Shu}  \email{panshu1987@163.com}
\affiliation{School of Sciences, Xi'an University of Technology - Xi'an 710054, China}

\author{Hui Gao} 
\affiliation{Web Sciences Center, University of Electronic
Science and Technology of China, Chengdu 610054, China}
\affiliation{Big Data Research Center, University of Electronic
Science and Technology of China, Chengdu 610054, China}

\author{Lidia A. Braunstein}
\affiliation{Center for Polymer Studies and Department of Physics,
Boston University, Boston, Massachusetts 02215, USA}
\affiliation{Instituto de Investigaciones F\'{i}sicas de Mar del
Plata (IFIMAR)-Departamento de F\'{i}sica,
Facultad de Ciencias Exactas y Naturales,
Universidad Nacional de Mar del Plata-CONICET,
Funes 3350, (7600) Mar del Plata, Argentina.}

\date{\today}
\begin{abstract}

\noindent
Promoting information spreading is a booming research topic
  in network science community. However, the exiting studies about
  promoting information spreading seldom took into account the human
  memory, which plays an important role in the spreading dynamics. In
  this paper we propose a non-Markovian information spreading model
  on complex networks, in which every informed node contacts a
  neighbor by using the memory of neighbor's accumulated contact
  numbers in the past. We systematically study the information
  spreading dynamics on uncorrelated configuration networks and a group
  of $22$ real-world networks, and find an effective contact strategy
  of promoting information spreading, i.e., the informed nodes
  preferentially contact neighbors with small number of accumulated
  contacts. According to the effective contact strategy, the high
  degree nodes are more likely to be chosen as the contacted neighbors
  in the early stage of the spreading, while in the late stage of the
  dynamics, the nodes with small degrees are preferentially
  contacted. We also propose a mean-field theory to describe our
  model, which qualitatively agrees well with the stochastic
  simulations on both artificial and real-world networks.
\end{abstract}


\maketitle

\section{Introduction}
A wide range of propagation phenomenon in the real world, such as the
spreading of information, rumor, disease and behavior, can be
described by spreading dynamics on complex
networks~\cite{pastor2015epidemic,wang2016unification,pastor2001epidemic,may2001infection,wang2016statistical,wang2014spatial}. The
problem of how to enhance or promote the spreading has attracted much
attention in the last few years in many
disciplines~\cite{pei2013spreading}. Promoting spreading speed and
understanding its effects in the outbreak size of the reached nodes
are two important features to study in both theoretical and empirical
aspects. Theoretically existing studies found that promoting the
spreading dynamics could induce distinct critical phenomena with
different outbreak thresholds and critical
exponents~\cite{ghoshal2011ranking}. Practically speaking, promoting
the spreading dynamics could shed some lights on the propagation of
information~\cite{iribarren2009impact,miritello2011dynamical,hambrick2012six},
marketing management~\cite{watts2007influentials,goel2010predicting},
disease
spreading~\cite{granell2013dynamical,wang2014asymmetrically,buono2013slow,buono2015immunization,alvarez2016interacting},
etc. Many strategies for promoting spreading dynamics have been
proposed, such as choosing influential
seeds~\cite{morone2015influence,kitsak2010identification} and
designing effective contact
strategies~\cite{yang2008selectivity,gao2016effective}. Kitsak
\emph{et~al.} found that selecting nodes with high \emph{k}-shells as
spreading sources can effectively enhance the size of the spreading in
most cases~\cite{kitsak2010identification}, however the \emph{k}-shell
index cannot reflect the importance of nodes in the core-like group in
which the nodes are connected very locally~\cite{liu2015core}.  Yang
\emph{et~al.} proposed a contact strategy based on the degree of
neighbors nodes to promote the information
spreading~\cite{yang2008selectivity}. They found that the information
spreading can be greatly promoted in uncorrelated networks if the
small-degree neighbors are preferentially contacted. In addition, if
the reached nodes, denoted as informed nodes, preferentially contact
nodes with few informed neighbors, the information spreading could be
further promoted~\cite{gao2016effective}.

In real-world, the activities of humans exhibit the characteristic of
having memory~\cite{barabasi2005origin,zhou2012relative}. This
characteristic has significant impacts on the spreading dynamics of
information, epidemic, and
behavior~\cite{vazquez2007impact0,wang2015dynamics,yang2011impact}.
For instance, human's memory produces a larger prevalence in the
exponential decay time of new informed nodes than processes without
memory~\cite{vazquez2007impact}. Wang
\emph{et~al.}~\cite{wang2015dynamics} found that memory affects the
growth pattern of the final outbreak size in the dynamics of social
contagion. However, previous works about promoting information
spreading always neglected the memory of individuals. In this paper,
we propose a non-Markovian information spreading model, in which each
individual (node) keeps memory of the number of accumulated contact
(NAC) with informed neighbors in the past. We assume that every
informed node contacts a neighbor based on the values of neighbors'
NACs. To describe our model, we develop a novel mean-field theory. Our
theoretical predictions are in good qualitative agreement with the
stochastic simulations on both artificial networks and $22$ real-world
networks. Through theoretical analysis and extensive numerical
simulations, we find an efficient strategy to promote the spreading
which consists on preferentially contacting neighbors with small
NACs. This strategy markedly promotes the information spreading, since
it increases the number of effective contacts with susceptible
nodes. With our effective strategy, we find that the informed nodes
are more likely to contact high degree nodes in the early stage, while
small degree nodes are preferentially contacted in the late stage. As
a result, our strategy unifies the probability that nodes of
different degrees will be contacted enhancing the
information spreading remarkably.

\section{Model}

We propose a generalized susceptible-informed
model~\cite{pastor2015epidemic} to describe the information
spreading. In this model, each node can either be in the susceptible
or informed state. Initially, we randomly choose a small fraction of
nodes in the informed state, while the remaining nodes are in the
susceptible state. At each time step $t$, each informed node $i$
contacts one of its neighbors $j$ with a probability $w_{ij}(t)$ (to
be defined later). If node $j$ is susceptible, it will become informed
with a transmission probability $\lambda$.  During the spreading
process, we adopt the synchronous updating rule, i.e., all nodes will
update their states synchronously at each time
step~\cite{shu2016recovery}. The dynamical process evolves until time
$T$, at which we compute the density of informed nodes $\rho(T)$. The
change of this magnitude with the parameters will allow us to evaluate
the efficiency of our strategy~\cite{liu2016locating}.

\begin{figure}[!htbp]
\centerline{\includegraphics[width=0.8\linewidth]{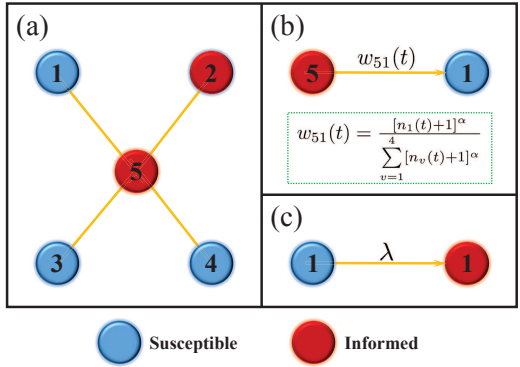}}
\caption{An illustration of the information spreading process on
  complex networks. (a) At time $t$, the informed node $5$ will
  contact a neighbor and transmit the information to it if the
  contacted neighbor is in the susceptible state. (b) Individual $5$
  contacts neighbor $j\in\{1,2,3,4\}$ with probability
  $w_{5j}(t)=[n_j(t)+1]^\alpha/\sum_{v=1}^4[n_v(t)+1]^\alpha$, where
  $n_j(t)$ is the NAC of node $j$. For example, node $5$ contacts
  node $1$ with probability $w_{51}(t)$. (c) Assuming that node $5$
  contacts node $1$ at time $t$ successfully, node $1$ will get the
  information and change to the informed state with probability
  $\lambda$.}\label{fig1}
\end{figure}

An effective information spreading strategy should increase the
effective contacts between informed and susceptible
nodes~\cite{toyoizumi2012reverse}. To achieve this goal our strategy
assumes that every node remembers the number of accumulated contact
(NAC) with informed neighbors in the past. An informed node $i$ contacts
a neighbor $j$ at time $t$ with probability
\begin{equation}\label{eq1}
w_{ij}(t)=\frac{[n_j(t)+1]^\alpha}{\sum\limits_{v\in \Gamma(i)}[n_v(t)+1]^\alpha},
\end{equation}
where $n_j(t)$ is the value of NAC of node $j$ at time $t$, and
$\Gamma_i$ is the neighbor set of node $i$~\cite{note1}. The
parameter $\alpha$ determines the tendency of node $i$ to contact a
neighbor $j$ with small or large value of $n_j(t)$. For the case of
$\alpha>0$, neighbors with larger $n_j(t)$ are preferentially
contacted, and for the case of $\alpha<0$, the opposite situation
happens. When $\alpha=0$, all neighbors are randomly chosen and we
recover the classical susceptible-informed model. As every node
remembers its NAC before time $t$, the information spreading process
is a non-Markovian process. In Figs.~\ref{fig1} we show a schematic of
the information spreading dynamics.

\section{Results}

\subsection{Theoretical analysis}
To describe our model of information spreading dynamics, we develop a
mean-field theory. We denote as $s_i(t)$ and $\rho_i(t)$ to the
probabilities that node $i$ is susceptible and informed at time $t$
respectively. Since each node can only be in susceptible or informed
state, we have $s_i(t)=1-\rho_i(t)$. A susceptible node $j$ will
become informed at time $t$ if it fulfills two conditions simultaneously
: (1) being contacted by an informed neighbor $i$ with
probability $w_{ij}(t)$, and (2) being informed successfully with
probability $\lambda$.  The probability of node $j$ to make a
transition to the informed state by neighbors at time $t$
is~\cite{gomez2010discrete}
\begin{equation}\label{eq2}
\Phi_j(t)=1-\prod\limits_{i=1}^N [1-\lambda
  A_{ij}w_{ij}(t)\rho_{i}(t)],
\end{equation}
where $N$ is the system size and $A_{ij}$ is the element
of the adjacency matrix of a given network. If there is an edge
between nodes $i$ and $j$, $A_{ij}=1$; otherwise, $A_{ij}=0$.
Thus the evolution of the probability of informed node $j$ is
\begin{equation}\label{eq3}
\rho_{j}(t+1)=\rho_{j}(t)+ s_{j}(t) \Phi_j(t).
\end{equation}
In order to obtain the value of $w_{ij}(t)$, we compute $n_j(t)$ which
is given by
\begin{equation}\label{eq4}
n_j(t+1)=n_j(t)+\Theta_j(t),
\end{equation}
where
\begin{equation}\label{eq5}
\Theta_j(t)= \sum\limits_{i=1}^NA_{ij}w_{ij}(t)\rho_{i}(t)
\end{equation}
is the increase of node $j$'s NAC at time $t$.
From Eqs.~(\ref{eq2})-(\ref{eq5}), we obtain the evolution equations
for the information spreading dynamics. At time $t$, the density of
informed nodes is given by
\begin{equation}\label{eq6}
\rho(t)=\frac{1}{N}\sum\limits_{j=1}^N \rho_{j}(t),
\end{equation}
and the density of susceptible nodes is $s(t)=1-\rho(t)$.

\subsection{Stochastic Simulations}

We perform extensive numerical simulations on both artificial and
real-world networks. All the obtained results are averaged over $10^3$
independent realizations of seeds on a fixed network.  Initially, we
randomly select 0.005 nodes as informed seeds.  The
information transmission probability is set as $\lambda=0.1$.  Notice
that other values of $\lambda > 0$ do not qualitatively affect our
results.

We first study the information spreading dynamics on artificial
networks. We built the artificial networks using the uncorrelated
configuration model~\cite{molloy1998size,catanzaro2005generation} with
power-law degree distributions $P(k)\sim k^{-\gamma}$ with
$k_{min}\leq k\leq\sqrt{N}$, where $k_{min}$ is the smaller value of
the degree and $\gamma$ is the degree exponent. In all our simulations
the network size is $N=10^4$ and $k_{min}=4$ in order to have a high
average degree and therefore a relative high number of contacts.
\begin{figure}[!htbp]
\centerline{\includegraphics[width=1\linewidth]{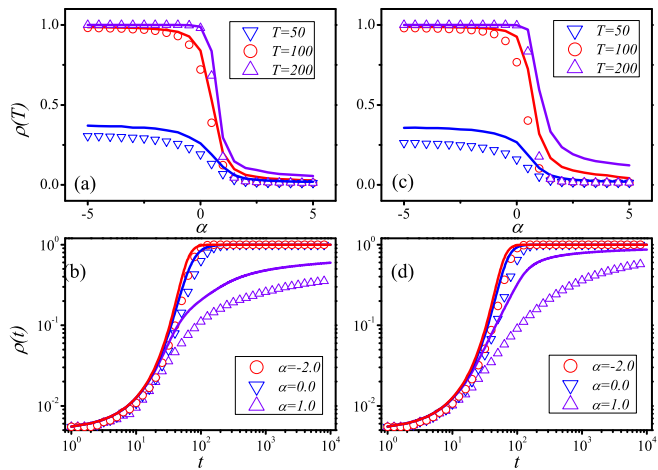}}
\caption{Information spreading on artificial networks with different
  degree exponents. The fractions of informed nodes at time $T=50,100$
  and $200$ as a function of $\alpha$ for $\gamma=2.1$ (a) and
  $\gamma=3.0$ (c). Time evolution of $\rho(t)$ in log-log scale with
  $\alpha=-2.0,0.0$ and $1.0$ versus $t$ for $\gamma=2.1$ (b) and
  $\gamma=3.0$ (d). The symbols represent the simulation results, and
  the lines are the theoretical results from
  Eq.~(\ref{eq6}).}\label{fig2}
\end{figure}

In Fig.~\ref{fig2} we show the results on artificial networks with
different degree exponents. For a given value of $T$, we find that
$\rho(T)$ decreases with $\alpha$, as shown in Figs.~\ref{fig2}(a) and
(c). Specifically, if every informed node $i$ preferentially contacts
a neighbor $j$ with small value of $n_j(t)$, i.e., $\alpha<0$, more
nodes will be informed. On the other hand, if neighbors with large
NACs are preferentially contacted, i.e., $\alpha > 0$, there will be
few nodes to be informed. From the figures we can see that for $T=100$
almost all nodes are in the informed state when $\alpha\leq-1$,
however, there are few informed nodes when $\alpha\geq 1$. We can
explain the above phenomena as follows. The larger the value of
$n_j(t)$, the larger the probability that node $j$ has been
informed. As a result, an informed node $i$ should contact a neighbor
with small value of NAC to increase the number of effective contacts
(i.e., contact with a susceptible neighbor), and further promotes the
information spreading. In Figs.~\ref{fig2} (b) and (d) we show the
time evolution of $\rho(t)$ for $\gamma=2.1$ and $3.0$,
respectively. We find that for small values of $\alpha$, $\rho(t)$ is
large, nevertheless when $\alpha=1.0$ the information spreads slowly
and it is hard to expand the information to the whole network. Our
theoretical predictions agree with the stochastic simulations in
most cases. The deviations between the theoretical and the
numerical results arise from disregarding the strong dynamical
correlations among the states of neighbors in the
theory~\cite{wang2016unification,pastor2015epidemic}. From the figures
we can also see that the phenomena for $\gamma=2.1$ and $\gamma=3.0$ are similar, which indicates that the heterogeneity of the degree distribution does
not qualitatively affect the results.

\begin{figure}[htbp]
\centerline{\includegraphics[width=1\linewidth]{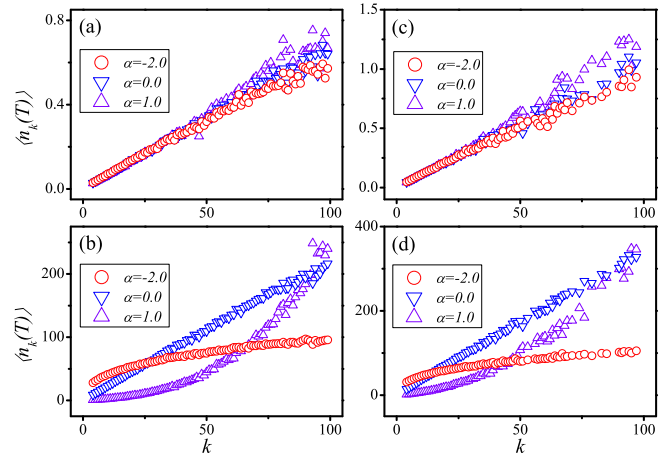}}
\caption{The average number of accumulated contacts $\langle n_k(T)\rangle$ for nodes with degree $k$ at time $T=10$ (top) and
  $T=100$ (bottom).  In (a) and (b), the degree exponent is
  $\gamma=2.1$.  In (c) and (d), we set $\gamma=3.0$.  }\label{fig3}
\end{figure}

In order to explain our above results we compute the average values of
NAC of nodes with degree $k$ $\langle n_k(T)\rangle$ at time $T$. From
Figs.~\ref{fig3}(a) and (c), we can observe that at the early stage of
the information spreading dynamics (i.e., $T=10$), $\langle
n_k(T)\rangle$ increases linearly with $k$ for all the values of
$\alpha$, i.e., large degree nodes are more likely to be contacted by
informed neighbors than low degree nodes. Note that $\langle
n_k(T)\rangle$ is slightly smaller for $\alpha=-2.0$ than for
$\alpha=1.0$. At later stage of the information spreading dynamics
(i.e., $T=100$), we find some interesting phenomena as shown in
Figs.~\ref{fig3}(b) and (d).  For the case of $\alpha=0.0$, $\langle
n_k(T)\rangle$ still increases linearly with $k$. When $\alpha=1.0$,
the values of $\langle n_k(T)\rangle$ of the largest degree nodes are
more than $200$ times larger than the ones for nodes with minimum
degree, and the number of the effective contacts are decreased
in this case. However, for the
case of $\alpha=-2.0$, almost all the nodes have the same values of
$\langle n_k(T)\rangle$. This indicates that, compared to the case of
$\alpha=1.0$, in the latter case informed nodes are more likely to
contact neighbors with small degrees. Since nodes with small degrees
have small probabilities to be informed, preferentially contacting
them increases the number of effective contacts. As a result, nodes
with different degrees almost have uniform probabilities of being
contacted and the information spreading can be accelerated
significantly.

\begin{figure}[htbp]
\centerline{\includegraphics[width=1\linewidth]{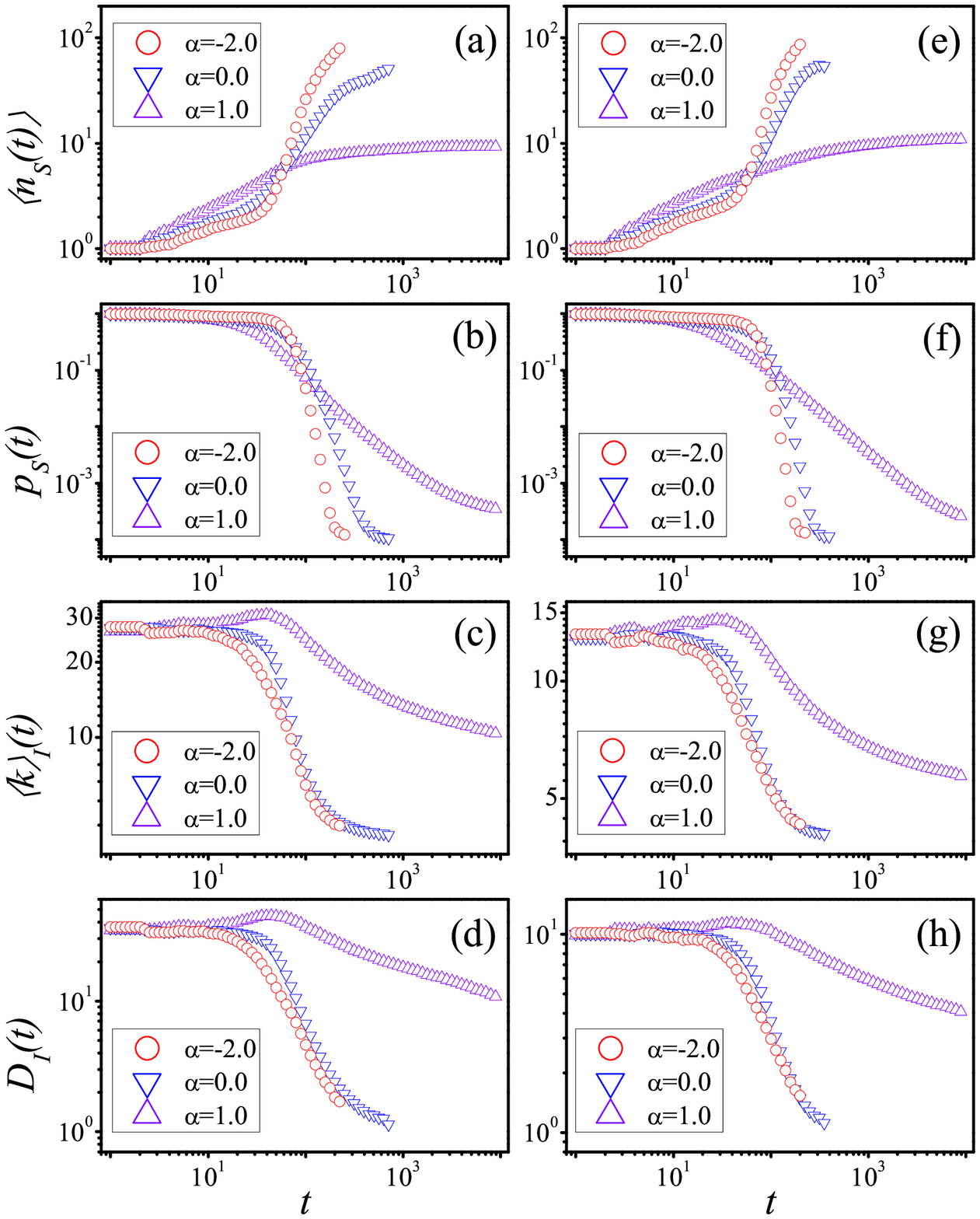}}
\caption{The time evolution properties of the information
spreading on theoretical networks. $\langle n_S(t)\rangle$
[(a) and (e)], $p_S(t)$ [(b) and (f)], $\langle k_I(t)
\rangle$ [(c) and (g)], and $D_I(t)$ [(d) and (h)] versus $t$.
In the left column, we set $\gamma=2.1$. In the right
column, we set $\gamma=3.0$. The information transmission
probability is $\lambda=0.1$.}\label{fig4}
\end{figure}

In Fig.~\ref{fig4}, we study the time evolution properties of the
information spreading. In Figs.~\ref{fig4}(a) and (e), we show the
time evolution of the average number of accumulated contacts $\langle
n_S(t)\rangle$ for the effective contacts (i.e., the newly contacted
susceptible nodes) for $\gamma=2.1$ and $3.0$. For the case of
$\alpha=-2.0$, most nodes with small number of contacts are informed
at the early stage, while the remaining few nodes with large number of
contacts are informed at the late stage. However, for the case of
$\alpha=1.0$, nodes with small number of contacts are hard be
informed. We also find that the ratio
of the number of effective contacts to all contacts at time
$t$, i.e., $p_S(t)$, is very high in the early stage when
$\alpha=-2.0$ [see Figs.~\ref{fig4}(b) and (f)]. To clarify the types
of informed nodes at different stages of the process, we study the
average degree $\langle k_I(t) \rangle$ and the degree diversity
$D_I(t)$~\cite{gao2016effective} of the newly informed nodes at time
$t$. Here $D_I(t)$ is defined as
\begin{equation}\label{eq8}
D_I(t)=\frac{1}{\sum\limits_{k}\left[\frac{I_k(t)-I_k(t-1)}{I(t)-I(t-1)}\right]^{2}},
\end{equation}
where $I(t)$ is the number of newly informed nodes at time $t$, and
$I_k(t)$ is the number of those with degree $k$. The large values of
$D_I(t)$ indicate that the newly informed nodes have heterogeneous
degrees.  From Figs.~\ref{fig4}(c) and (g), we can see that for the
case of $\alpha=-2.0$, $\langle k_I(t) \rangle$ is very large at the
early stage, and then decreases at the later stage. $D_I(t)$ shows the
same tendency as seen in Figs.~\ref{fig4}(d) and (h). Thus, when
$\alpha=-2.0$ high degree nodes are more likely to be informed in the
early stage, while in the late stage small degree nodes are often
informed. When $\alpha=1.0$ both $\langle k_I(t) \rangle$ and $D_I(t)$
are large in the whole spreading process, which means that nodes with
small degrees are hard to be informed. These results corroborate our
finding that the heterogeneity of degree distribution does not
qualitatively affect the above stated results.

Finally, we study our suggest information spreading dynamics on $22$
real-world networks, which includes metabolic networks, infrastructure
networks, collaboration networks and citation networks, etc. For
simplicity, the directed networks are treated as undirected ones and
the weighted networks are treated as unweighted
ones. Table~\ref{table_Ch} displays their statistical characteristics
in details.

\begin{figure}[htbp]
\centerline{\includegraphics[width=1\linewidth]{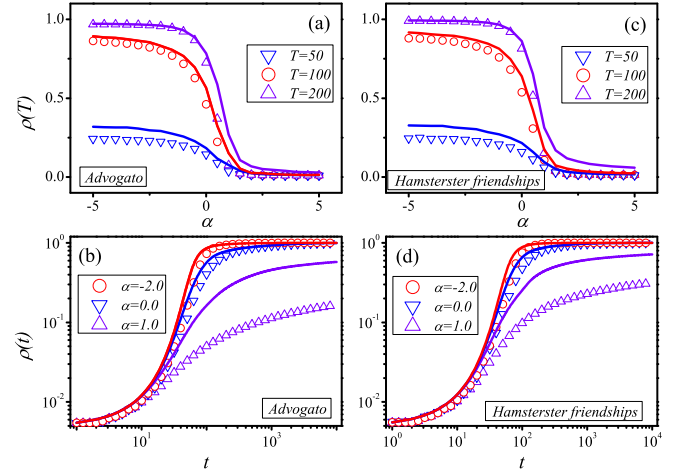}}
\caption{Information spreading on real-world networks. The fractions
  of informed nodes at time $T=50,100$ and $200$ for Advogato (a) and
  Hamsterster friendships networks (c). The time evolution of
  $\rho(t)$ with $\alpha=-2.0,0.0$ and $1.0$ versus $t$ for Advogato
  (c) and Hamsterster friendships networks (d). The theoretical
  analysis results are obtained from Eq.~(\ref{eq6}). We set the
  information transmission probability as $\lambda=0.1$.}\label{fig5}
\end{figure}

Fig.~\ref{fig5} shows the information spreading on two representative
networks. Strikingly, we find that the results on real-world networks
are qualitatively similar to the one found in artificial networks as
shown in Fig.~\ref{fig2}. Specifically, if the informed nodes
preferentially contact neighbors $j$ with small $n_j(t)$, the
information spreading will be markedly facilitated. On the contrary,
the information spreads is slower if nodes with large NACs are
favored, and it is hard to spread the information to the network.

\begin{figure}[htbp]
\centerline{\includegraphics[width=0.7\linewidth]{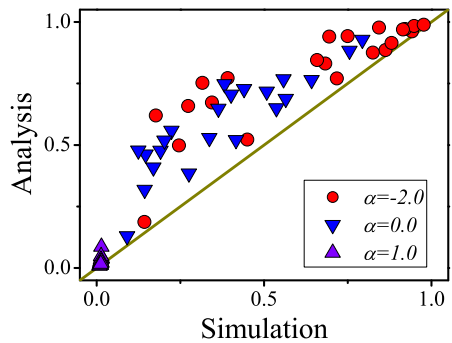}}
\caption{Comparing the theoretical and numerical predictions of the
  information spreading on $22$ real-world networks. Different colors
  indicate different values of $\alpha$. The analysis results are
  obtained from Eq.~(\ref{eq6}). The results are obtained at the end
  time $T=100$. We set $\lambda=0.1$.}\label{fig6}
\end{figure}

We further verify the effectiveness of our suggested mean-field theory
on the real-world networks at $T=100$ for different values of
$\alpha$, as shown in Fig.~\ref{fig6}.  We can see that our
theoretical predictions qualitatively agree with the numerical
simulations, although the theoretical predictions are slightly larger than the simulation
results. The deviations between theoretical and numerical predictions
derive from the strong dynamical correlations among the states of
neighbors~\cite{wang2016unification,pastor2015epidemic}.

\begin{table*}
\scriptsize
\caption{Statistical characteristics of the $22$ real-world
  networks. The statistical characteristics including the network size
  ($N$), number of edges ($E$), maximum degree ($k_{max}$), first and
  second moments of the degree distribution ($\langle k\rangle$) and
  second moments ($\langle k^2\rangle$), degree-degree correlations
  ($r$), clustering ($c$), and modularity ($Q$). } \label{table_Ch}
\centering
  \begin{tabular}{ l| c |c c c c c c c c }
    \hline
    \hline
    \multirow{2}{*}{\textbf{Category}} &
    \multirow{2}{*}{\textbf{Networks}} & \multicolumn{8}{ c}{\textbf{Statistical Characteristics of Networks}}\\
    \cline{3-10}
      &   & $N$ & $E$ & $k_{max}$ & $\langle k\rangle$ &  $\langle k^2\rangle$ & $r$ & $c$ & $Q$ \\
      \hline

    \multirow{2}{*}{Social} &Advogato~\cite{massa2009bowling}&  5042& 39227&  803&15.56&1284&$-0.096$&0.092&0.337\\
       & Hamsterster friendships~\cite{http}  & 1788& 12476&  272&13.955&635.61&$-0.089$&0.09&0.396\\
       &Hamsterster full~\cite{http}  & 2000& 16098&   273&16.098&704.71&0.023&0.23&0.45\\
    \hline
    \multirow{4}{*}{Metabolic}
      &Caenorhabditis elegans~\cite{duch2005community}  & 453& 2025&  237&8.9404&358.49&$-0.226$&0.124&0.401\\
       &Reactome~\cite{joshi2005reactome}  &  5973 & 145778 &  855&48.812&6995.1&0.241&0.606&0.719\\
       &Human protein (Figeys)~\cite{ewing2007large} &  2217& 6418&  314&5.79&324.93&$-0.332$&0.008&0.472\\
            &Human protein (Vidal)~\cite{rual2005towards} & 2783& 6007&   129&4.317&68.103&$-0.137$&0.035&0.615\\
          \hline
          \multirow{3}{*}{Infrastructure}
       &Air traffic control~\cite{http}   & 1226& 2408&  34&3.928&28.899&$-0.015$&0.064&0.686\\
       &OpenFlights~\cite{opsahl2010node}   & 2905& 15645&  242&10.771&601.45&0.049&0.255&0.581\\
       &Power~\cite{watts1998collective}   & 4941& 6594& 19&2.669&10.333&0.003&0.08&0.932\\
     \hline
      \multirow{2}{*}{Collaboration}
       &CA-Hep~\cite{leskovec2007graph}  & 8638& 24806&  65&5.744&74.601&0.239&0.482&0.752\\
       &Netsci~\cite{newman2006finding}  &379& 914& 34&4.823&38.686&$-0.082$&0.741&0.846\\
    \hline
     \multirow{1}{*}{Citation}
       &DBLP~\cite{ley2002dblp}  & 12495& 49563&709&7.933&347.28&$-0.046$&0.062&0.538\\
     \hline
     \multirow{2}{*}{HumanSocial}
       & Jazz musicians~\cite{gleiser2003community} & 198& 2742&   100&27.697&1070.2&0.02&0.52&0.439\\
      &Adolescent health~\cite{moody2001peer} & 2539& 10455&  27&8.236&86.414&0.251&0.142&0.597\\
     \hline
     \multirow{1}{*}{Computer}
      &Route views~\cite{leskovec2007dynamics}  & 6474& 12572&  1458&3.884&640.08&$-0.182$&0.01&0.612\\
      &Route~\cite{spring2004measuring}  & 5022& 6258&  106&2.492&34.181&$-0.138$&0.012&0.898\\
     \hline
     \multirow{1}{*}{Lexical}
      & King James~\cite{http}& 1707& 9059&  364&10.614&441.85&$-0.052$&0.162&0.461\\

     \hline
     \multirow{1}{*}{OnlineContact}
     &Pretty Good Privacy~\cite{boguna2004models}  & 10680& 24340&  206&4.558&86.287&0.239&0.266&0.877\\
     &Email~\cite{guimera2003self}  & 1133& 5451&71&9.622&179.18&0.078&0.22&0.565\\
     &Emailcontact~\cite{kitsak2010identification}  &12625& 20362&576&3.226&356.36&$-0.387$&0.109&0.674\\
     &Blog~\cite{xie2006social}  & 3982& 6803&189&3.417&47.145&$-0.133$&0.284&0.853\\
     \hline
    \hline
  \end{tabular}
\end{table*}

\section{Discussions}

In this paper, we proposed a novel non-Markovian information
spreading model by introducing the individuals' memory, which assumes
that every node remembers the number of accumulated contacts (NAC)
with informed neighbors in the past, and an informed node contacts a
neighbor with different biases values of NAC. To describe the
non-Markovian spreading dynamics, we developed a novel mean-field
theory. Through extensive numerical simulations on artificial networks
and real-world networks, we found that the memory characteristic
markedly affects the spreading dynamics. If the informed nodes
preferentially contact neighbors with low NACs, the information
spreading is markedly promoted due to the increasing of the number of effective contacts between susceptible and informed nodes. In the effective contact strategy, the high degree nodes are
more likely to be contacted in the early stage, while the small degree
nodes are preferentially contacted in the late stage. As a result,
nodes with different degrees almost have uniform probabilities of
being contacted and the information spreading is accelerated
significantly. Importantly, the effectiveness of the promoting
strategy weakly depends on the structures of networks. The agreements
between our theoretical predictions and numerical results are verified
on artificial networks as well as the real-world networks. Our
effective strategy to promote the information spreading on complex
networks, could be used for other spreading dynamics, such as
technical innovations, healthy behaviors, and new products
adoption. Our theoretical method of non-Markovian spreading model can
be extended to the study of other social dynamics.

\acknowledgments This work was supported by the National Natural
Science Foundation of China (Grant
No. 61673086). LAB thanks UNMDP and
  FONCyT, Pict 0429/2013 for financial support.


\begin{thebibliography}{54}
\bibitem[{not()}]{note1}
\bibinfo{note}{In the equation, we set $n_j(t)+1$ to avoid the denominator is
  zero.}
\end{thebibliography}

\end{document}